\documentclass[preprintnumbers,superscriptaddress,pra,showkeys]{revtex4}
\usepackage{amsfonts}
\usepackage{amssymb}
\usepackage{amsmath}
\usepackage{epsfig}
\usepackage{graphicx}
\usepackage{color}

\input{tcilatex}
\begin{document}

\title{Local shape of the vapor-liquid critical point on the thermodynamic
surface and the van der Waals equation of state}
\author{J. S. Yu}
\affiliation{School for Theoretical Physics, School of Physics and Electronics, Hunan
University, Changsha 410082, China}
\author{X. Zhou}
\affiliation{School for Theoretical Physics, School of Physics and Electronics, Hunan
University, Changsha 410082, China}
\author{J. F. Chen}
\affiliation{School for Theoretical Physics, School of Physics and Electronics, Hunan
University, Changsha 410082, China}
\author{W. K. Du}
\affiliation{School for Theoretical Physics, School of Physics and Electronics, Hunan
University, Changsha 410082, China}
\author{X. Wang}
\affiliation{School for Theoretical Physics, School of Physics and Electronics, Hunan
University, Changsha 410082, China}
\author{Q. H. Liu}
\email{quanhuiliu@gmail.com}
\affiliation{School for Theoretical Physics, School of Physics and Electronics, Hunan
University, Changsha 410082, China}
\date{\today }

\begin{abstract}
Differential geometry is powerful tool to analyze the vapor-liquid critical
point on the surface of the thermodynamic equation of state. The existence
of usual condition of the critical point $\left( \partial p/\partial
V\right) _{T}=0$ requires the isothermal process, but the universality of
the critical point is its independence of whatever process is taken, and so
we can assume $\left( \partial p/\partial T\right) _{V}=0$. The distinction
between the critical point and other points on the surface leads us to
further assume that the critical point is geometrically represented by zero
Gaussian curvature. A slight extension of the van der Waals equation of
state is to letting two parameters $a$ and $b$ in it vary with temperature,
which then satisfies both assumptions and reproduces its usual form when the
temperature is approximately the critical one. 
\end{abstract}

\author{}
\maketitle

\section{Introduction}

In thermal physics, a critical point is the end point of a phase equilibrium
curve, the pressure--temperature curve that designates conditions under
which a liquid phase and a vapor phase can coexist. The critical point ($%
T_{C},V_{C},p_{C}$) in the\ $pV$diagrams determined by,%
\begin{equation}
\left( \frac{\partial p}{\partial V}\right) _{T}=0,\left( \frac{\partial
^{2}p}{\partial V^{2}}\right) _{T}=0,  \label{cp}
\end{equation}%
together with the thermodynamic equation of state (EoS), where symbols ($%
T,V,p$) have their usual meaning in ordinary textbooks \cite%
{text1,text2,text3,text4,text5}. The phase transition exhibits critical
slowing down, universality and scaling, etc., which reflect a fact that the
details of the system play insignificant role \cite{lpk,wilson}. How to
characterize the essence of the critical point is always an attractive
topic. We note two seemly independent developments/facts. One is that the
critical slowing down is its path-independence \cite%
{csd1,csd2,csd3,csd4,csd5}, which means that starting from any thermodynamic
state in the vicinity of a critical point to approach to it, the system has
inherently slow timescales whatever thermodynamic processes are chosen. The
second is that a geometrical description of a local point on a curved
surface is irrespective of either the parameters chosen to label the point
of the surface or the paths selected to approach to it. The strong
resemblance of these two facts suggests that geometrical description of the
critical point is advantageous. Based on this observation, we make a
proposal that the critical point is geometrically represented by zero
Gaussian curvature on the thermodynamic EoS surface, together with some
physical assumptions. We hope to use this proposal to resolve a long
standing problem associated with the van der Waals (vdW) EoS.

The most prominent aspect of the vdW EoS is that it captures many of the
qualitative features of the liquid--vapor phase transition with possible
help of Maxwell's equal area rule. The vdW EoS was\ essentially presented 
\cite{history1} (but explicitly given later \cite{history2}) by van der
Waals in his 1873 Ph. D. thesis, and for this he was awarded the Nobel Prize
in Physics 1910 \cite{history1,history2,history3,history4,review}. The vdW
EoS is well-known,%
\begin{equation}
p=\frac{nRT}{V-nb}-\frac{n^{2}a}{V^{2}},  \label{vdw}
\end{equation}%
where two parameters, $a$ and $b$, can be estimated from the critical point
and considered constants which are specific for each substance, and other
symbols ($n,R$) also have their usual meaning in ordinary textbooks \cite%
{text1,text2,text3,text4,text5}. For one mole fluid $n=1$, and the values of 
$T_{C},V_{C},p_{C}$ are in terms of $a$ and $b$ parameter \cite%
{text1,text2,text3,text4,text5},%
\begin{equation}
T_{C}=\frac{8a}{27Rb},p_{C}=\frac{a}{27b^{2}},V_{C}=3b.  \label{C}
\end{equation}%
With these values, the vdW EoS can be transformed into following
dimensionless form,%
\begin{equation}
p^{\ast }=\frac{8}{3}\frac{t^{\ast }}{v^{\ast }-1/3}-\frac{3}{v^{\ast 2}},
\label{csl}
\end{equation}%
where,%
\begin{equation}
t^{\ast }\equiv \frac{T}{T_{C}},\text{ }v^{\ast }\equiv \frac{V}{V_{C}}%
,p^{\ast }\equiv \frac{p}{p_{C}}.  \label{tvp}
\end{equation}%
The equation (\ref{csl}) is referred as the Law of Corresponding States
which holds for all kinds of fluid substances, which was also originated
with the work of van der Waals in about 1873 \cite{history1}, when he used
the critical temperature and critical pressure to characterize a fluid.
However, whether and how the vdW parameters $a$ and $b$ depend on the
temperature $T$, and even more, on the volume $V$, has been a problem of
long history. van der Waals himself was well-aware of it\ \cite{history2},
and remarked in his Nobel prize speech: "\textit{I have never been able to
consider that the last word had been said about the equation of state and I
have continually returned to it during other studies. As early as 1873 I
recognized the possibility that }$a$\textit{\ and }$b$\textit{\ might vary
with temperature, and it is well-known that Clausius even assumed the value
of }$a$\textit{\ to be inversely proportional to the absolute temperature.}" 
\cite{history1} In fact, more than one century passed since the discovery of
the vdW EoS, we do not have a strong experimental evidence nor a compelling
theoretical argument to indicate how $a$ and $b$ parameter might depend on
the temperature and/or volume. We have some theoretical results in
statistical mechanics, revealing some temperature dependence of $a$ and $b$,
for instance in the hard-sphere model \cite{text1,text2,text3,text4,text5},
but these results are frequently obtained for dilute fluid far from the
critical point, and more importantly, they rely heavily on the specific
model without universality which is inherent to the thermodynamics.

The present paper thus addresses two problems. One is why we assume $\left(
\partial p/\partial T\right) _{V}=0$ that is complementary to the first
equation of (\ref{cp}),\ and why we propose that the critical point is
geometrically represented by zero Gaussian curvature on the thermodynamic
EoS surface. Another is to use above assumptions to discuss the
long-standing problem within the thermodynamics. The paper is organized in
the following. In section II, we prove a theorem stating that the local
shape of the vapor-liquid critical point on the thermodynamic surface can
never be an elliptic point; and in order to completely characterize the
local shape of the critical point, we need two more response functions which
are assumed to vanish at the critical point, implied by the critical slowing
down observed in either the realistic experiments or the computer simulation
of the phase transition \cite{csd1,csd2,csd3,csd4,csd5}. The vanishing
response functions leads to the zero Gaussian curvature. In section III, the
vdW EoS is slightly extended such that the parameters $a$ and $b$ vary with
the temperature $T$, which is thus capable to give\ zero Gaussian curvature
at the critical point, while the usual form of the vdW EoS fails. In section
IV, a brief summation of the present study is given.

In present paper, we concentrate the (interior) Gaussian curvature that is
sufficient to specify the local shape of the two-dimensional thermodynamic
EoS surface, but we will also give the (exterior)\ mean curvature as a
contrasting quantity. In geometry, the curvature is usually referred to the
interior one.

\section{Local shape of the vapor-liquid critical point on the EoS surface
and a proposal}

In differential geometry, the local shapes of a two-dimensional curved
surface are completely classified into three types:\ elliptic, hyperbolic
and parabolic, corresponding to the Gaussian curvature greater than, smaller
than or equal to, zero, respectively \cite{docarmo}. For a thermodynamic EoS 
$p=p(T,V)$ that can be treated as a two-dimensional surface in the
three-dimensional flat space of coordinates $p$, $T$ and $V$, we now show
that the vapor-liquid critical point can not be an elliptic point.

In geometry, it is preferable to use the dimensionless equation of the EoS
surface $p=p(T,V)$. The straightforward calculations can, respectively, give 
$H$ and Gaussian curvature $K$, 
\begin{eqnarray}
H &=&\frac{\left( \frac{\partial ^{2}p}{\partial V^{2}}\right) _{T}\left(
\left( \frac{\partial p}{\partial T}\right) _{V}^{2}+1\right) +\left( \frac{%
\partial ^{2}p}{\partial T^{2}}\right) _{V}\left( \left( \frac{\partial p}{%
\partial V}\right) _{T}^{2}+1\right) -2\left( \frac{\partial p}{\partial V}%
\right) _{T}\left( \frac{\partial p}{\partial T}\right) _{V}\left( \frac{%
\partial ^{2}p}{\partial V\partial T}\right) }{2\left( \left( \frac{\partial
p}{\partial V}\right) _{T}^{2}+\left( \frac{\partial p}{\partial T}\right)
_{V}^{2}+1\right) ^{3/2}},  \label{GenH} \\
K &=&\frac{\left( \frac{\partial ^{2}p}{\partial V^{2}}\right) _{T}\left( 
\frac{\partial ^{2}p}{\partial T^{2}}\right) _{V}-\left( \frac{\partial ^{2}p%
}{\partial V\partial T}\right) ^{2}}{\left( \left( \frac{\partial p}{%
\partial V}\right) _{T}^{2}+\left( \frac{\partial p}{\partial T}\right)
_{V}^{2}+1\right) ^{2}}.  \label{GenK}
\end{eqnarray}%
At the critical point the conditions (\ref{cp})$\ $apply, we have the mean
curvature $H_{C}$ and Gaussian curvature $K_{C}$, respectively, 
\begin{equation}
H_{C}=\frac{\left( \frac{\partial ^{2}p}{\partial T^{2}}\right) _{V}}{%
2\left( \left( \frac{\partial p}{\partial T}\right) _{V}^{2}+1\right) ^{3/2}}%
,K_{C}=-\frac{\left( \frac{\partial ^{2}p}{\partial V\partial T}\right) ^{2}%
}{\left( \left( \frac{\partial p}{\partial T}\right) _{V}^{2}+1\right) ^{2}},
\label{CHK}
\end{equation}%
which shows that $K\leq 0$. Thus, we in fact prove a theorem that the local
shape of the vapor-liquid critical point on the thermodynamic surface can
never be an elliptic point.

To illustrate the mean and Gaussian curvature of the surface of the
thermodynamic EoS, let us first consider two simple systems. For an
incompressible liquid EoS: $V=const.$, which is a flat plane, both
curvatures are zero. The ideal gas EoS surface, $p=nRT/V$ which can be
rewritten as a dimensionless one $p^{\ast }v^{\ast }=t^{^{\ast }}$ with a
reference point ($p_{0}$, $V_{0}$, $T_{0}\left( =p_{0}V_{0}/\left( nR\right)
\right) $), where $t^{\ast }\equiv T/T_{0},$ $v^{\ast }\equiv
V/V_{0},p^{\ast }\equiv p/p_{0}$. The mean curvature $H$ and Gaussian
curvature $K$ are, respectively, 
\begin{equation}
H=\frac{t^{\ast }v^{\ast 3}}{\left( t^{^{\ast }2}+v^{\ast 2}+v^{\ast
4}\right) ^{3/2}},K=-\frac{v^{\ast 4}}{\left( t^{^{\ast }2}+v^{^{\ast
}2}+v^{\ast 4}\right) ^{2}}.  \label{HKig}
\end{equation}%
Since the Gaussian curvature $K<0$ is negative definite, every point on the
ideal gas EoS surface is saddle.

Now, we examine vdW EoS surface (\ref{vdw}) and it is preferable to use the
dimensionless form (\ref{csl}). The mean curvature $H$ and Gaussian
curvature $K$ are, respectively, 
\begin{equation}
H=9v^{\ast 5}(3v^{\ast }-1)^{3}\frac{F_{1}(t^{\ast },v^{\ast })}{\left(
F_{2}(t^{\ast },v^{\ast })\right) ^{3/2}},K=-\frac{576(3v^{\ast
}-1)^{4}v^{\ast 12}}{\left( F_{2}(t^{\ast },v^{\ast })\right) ^{2}},
\label{HK}
\end{equation}%
where, 
\begin{equation}
F_{1}(t^{\ast },v^{\ast })=8t^{\ast }v^{\ast 4}-27v^{\ast 3}+27v^{\ast
2}-73v^{\ast }+65,  \label{F1}
\end{equation}%
\begin{eqnarray}
F_{2}(t^{\ast },v^{\ast }) &=&576t^{\ast 2}v^{\ast 6}-2592t^{\ast }v^{\ast
5}+1728t^{\ast }v^{\ast 4}-288t^{\ast }v^{\ast 3}  \notag \\
&&+81v^{\ast 10}-108v^{\ast 9}+630v^{\ast 8}-396v^{\ast 7}+65v^{\ast 6} 
\notag \\
&&+2916v^{\ast 4}-3888v^{\ast 3}+1944v^{\ast 2}-432v^{\ast }+36.  \label{F2}
\end{eqnarray}%
At the critical point $(t^{\ast },v^{\ast })=(1,1)$, we have, respectively, 
\begin{equation}
H_{C}=0,\text{ }K_{C}=-\frac{36}{289}\approx -0.125.  \label{HKC}
\end{equation}%
The negative Gaussian curvature $K_{C}\approx -0.125$ indicates that the
critical point is a hyperbolic point, more precisely, a saddle point \cite%
{docarmo}.

A comparison of the Gaussian curvatures (\ref{HKig}) for ideal gas and (\ref%
{HKC}) for vdW EoS suggests that there is no qualitative difference in
between. It is a little bit odd, for we believe that a realistic EoS differs
from the ideal gas EoS in the qualitative sense, rather than a quantitative
one. By the critical point on the $PV$ diagram, we mean a stationary
inflection point in the constant-temperature line, critical isotherm, whose
location is determined by two equations in (\ref{cp}). However, if one
approaches this point from an isobaric process or an isovolumetric process,
or a more complicated process, we do not know whether such a point exhibits
the same singularity. Therefore we must seek for a general condition for
critical point, independent of thermodynamic paths.

At the local point of the thermodynamic EoS surface $p=p(T,V)$, the
tangential plane is spanned by two independent vectors ($dT,dV$). At the
critical point ($T_{C},V_{C}$), we have $\left( \partial p/\partial V\right)
_{T}=0$ (\ref{cp}) which means existence of a limit along isotherm. Such a
limit must exist irrespective along isotherm ($T=const.$) or along
isovolumetric line ($V=const.$), implying that we can further impose $\left(
\partial p/\partial T\right) _{V}=0$ at the critical point. Another
condition is inspired by the Gaussian curvature that is independent of the
detailed structure of matter, and the simplest assumption is $K_{C}=0$,
implying $\partial ^{2}p/\partial V\partial T=0$. In sum, we propose two
additional conditions for the critical point on the EoS surface $p=p(T,V)$,%
\begin{equation}
\left( \partial p/\partial T\right) _{V}=0\text{, }\partial ^{2}p/\partial
V\partial T=0.  \label{csd}
\end{equation}%
It is worthy of mentioning that, in contrast to the realistic experiments
which seem hard to measure these two response functions near the critical
point, the computer simulations are more feasible \cite{csd1,csd2,csd3,csd4}
which show that the critical slowing down is\ really an overall phenomenon
no matter what path is chosen to approach to the critical point.

\section{The proposal and temperature dependence of vdW parameters $a$ and $%
b $}

We are confident that the vdW EoS with constant parameters $a$ and $b$ is
not satisfactory for following two senses. The first is that the Gaussian
curvature at the critical point is $K_{C}\approx -0.125$ (\ref{HKC}) which
is not qualitatively different from other point except the limiting
situation. The second is that this value $K_{C}\approx -0.125$ manifestly
depends on special thermodynamic path, i.e., isotherm (\ref{cp}).
Fortunately, the vdW EoS can be adapted for removal of these weaknesses.

The slightest extension of the vdW EoS is to let two constants $a$ and $b$
in the vdW EoS (\ref{vdw}) depend on the temperature as $a\rightarrow a(T)$
and $b\rightarrow b(T)$. The critical values of $T_{C},V_{C},p_{C}$ are
entirely determined by $a(T_{C})$ and $b(T_{C})$, 
\begin{equation}
T_{C}=\frac{8a(T_{C})}{27Rb(T_{C})},p_{C}=\frac{a(T_{C})}{27b(T_{C})^{2}}%
,V_{C}=3b(T_{C}).  \label{ec}
\end{equation}%
With introduction of the dimensionless $\alpha \left( t^{\ast }\right) $ and 
$\beta \left( t^{\ast }\right) $ instead of $a(T)$ and $b(T)$ in the
following,%
\begin{equation}
\alpha \left( t^{\ast }\right) \equiv \frac{a(T)}{a(T_{C})}=\frac{a(t^{\ast
}T_{C})}{a(T_{C})},\beta \left( t^{\ast }\right) \equiv \frac{b(T)}{b(T_{C})}%
=\frac{b(t^{\ast }T_{C})}{b(T_{C})},
\end{equation}%
the Law of Corresponding States does not hold true any more except the
special case, $\alpha =const.$ and $\beta =const.$; and we have instead the
dimensionless extended vdW EoS, 
\begin{equation}
p^{\ast }=\frac{8}{3}\frac{t^{\ast }}{v^{\ast }-\beta \left( t^{\ast
}\right) /3}-\frac{3\alpha \left( t^{\ast }\right) }{v^{\ast 2}}.
\label{exvdw}
\end{equation}%
Near the critical point, we assume that $a(T)$ and $b(T)$ parameters take
the following forms, 
\begin{subequations}
\begin{eqnarray}
a\left( T\right) &\approx &a\left( T_{C}\right) +a^{\prime }\left(
T_{C}\right) (T-T_{C})+\frac{1}{2}a^{^{\prime \prime }}\left( T_{C}\right)
^{2}(T-T_{C})^{2},  \label{at} \\
b\left( T\right) &\approx &b\left( T_{C}\right) +b^{\prime }\left(
T_{C}\right) (T-T_{C})+\frac{1}{2}b^{^{\prime \prime }}\left( T_{C}\right)
^{2}(T-T_{C})^{2},  \label{bt}
\end{eqnarray}%
where, 
\end{subequations}
\begin{equation}
g^{\prime }=\frac{dg}{dT},g=a,b,a^{\prime },b^{\prime },....  \label{g'}
\end{equation}%
The relations between set $\left( \alpha ^{\prime }\left( T_{C}\right)
,\beta ^{\prime }\left( T_{C}\right) \right) $ and set $\left( a^{\prime
}\left( T_{C}\right) ,b^{\prime }\left( T_{C}\right) \right) $ are,%
\begin{equation}
\alpha ^{\prime }\left( t^{\ast }=1\right) =T_{C}\frac{a^{\prime }\left(
T_{C}\right) }{a\left( T_{C}\right) },\beta ^{\prime }\left( t^{\ast
}=1\right) =T_{C}\frac{b^{\prime }\left( T_{C}\right) }{b\left( T_{C}\right) 
},  \label{last}
\end{equation}%
where $\alpha ^{\prime }=d\alpha /dt^{\ast }$ and $\beta ^{\prime }=d\beta
/dt^{\ast }$, etc.

The mean curvature $H$ and Gaussian curvature $K$ of the dimensionless
extended vdW EoS surface have very long expressions of complicated
structure. However, the expressions for both\ $H$ and $K$ at the critical
point $\left( t^{\ast },v^{\ast },\alpha ,\beta \right) =(1,1,1,1)$ are
simply, 
\begin{equation}
H_{C}=\frac{2\beta ^{\prime }\left( \beta ^{\prime }+2\right) -3\alpha
^{\prime \prime }+2\beta ^{\prime \prime }}{2\left( G(t^{\ast },v^{\ast
})\right) ^{3/2}},K_{C}=-\frac{36(1-\alpha ^{\prime }+\beta ^{\prime })^{2}}{%
\left( G(t^{\ast },v^{\ast })\right) ^{2}},  \label{EHK}
\end{equation}%
where,%
\begin{equation}
G(t^{\ast },v^{\ast })=4\beta ^{\prime 2}+9\alpha ^{\prime 2}-12\alpha
^{\prime }\beta ^{\prime }+16\beta ^{\prime }-24\alpha ^{\prime }+17.
\label{G}
\end{equation}

\emph{The distinctive feature of the extended vdW EoS is that it contains
two possible local shapes at the critical point}: \emph{hyperbolic and
parabolic}, for $K_{C}\leq 0$. The case $K_{C}=0$ can be realized provided, 
\begin{equation}
1-\alpha ^{\prime }+\beta ^{\prime }=0.  \label{flat}
\end{equation}

To note that two response functions $\left( \partial p/\partial T\right)
_{V} $, and its partial derivative with respect to volume, $\left( \partial
^{2}p/\partial V\partial T\right) $, produce values at the critical point $%
\left( T_{C},V_{C},p_{C}\right) $, respectively, 
\begin{subequations}
\begin{eqnarray}
\left( \frac{\partial p}{\partial T}\right) _{V} &=&\frac{9RT_{C}b^{\prime
}(T_{C})-4a^{\prime }(T_{C})+18Rb(T_{C})}{36\left( b\left( T_{C}\right)
\right) ^{2}},  \label{exp1} \\
\frac{\partial ^{2}p}{\partial V\partial T} &=&-\frac{27RT_{C}b^{\prime
}(T_{C})-8a^{\prime }(T_{C})+27Rb(T_{C})}{108\left( b\left( T_{C}\right)
\right) ^{3}}.  \label{exp2}
\end{eqnarray}%
These two values are sufficient to completely fix two derivatives $\left(
a^{\prime }\left( T_{C}\right) ,b^{\prime }\left( T_{C}\right) \right) $,
given that the parameter $b\left( T_{C}\right) $ in (\ref{exp1})-(\ref{exp2}%
) is given by the magnitude of the molar critical volume via (\ref{ec}).

Now let us examine situations where both response functions in (\ref{csd})
vanish at the critical point. First, once the second response function
vanishes at the critical point, $\left( \partial ^{2}p/\partial V\partial
T\right) =0$, i.e., $27RT_{C}b^{\prime }(T_{C})-8a^{\prime
}(T_{C})+27Rb(T_{C})=0$ from (\ref{exp2}), we have from the relations (\ref%
{last}), 
\end{subequations}
\begin{equation}
27Rb\left( T_{C}\right) \beta ^{\prime }-8\alpha ^{\prime }a\left(
T_{C}\right) +27RT_{C}b(T_{C})=0.
\end{equation}%
which reproduces $1-\alpha ^{\prime }+\beta ^{\prime }=0$ (\ref{flat}) with $%
27T_{C}Rb\left( T_{C}\right) =8a(T_{C})$ (\ref{ec}). Secondly, once the
first response function in (\ref{csd}) vanishes at the critical point, $%
\left( \partial p/\partial T\right) _{V}=0$, i.e., $9RT_{C}b^{\prime
}(T_{C})-4a^{\prime }(T_{C})+18Rb(T_{C})=0$ from (\ref{exp1}), we have $%
4-3\alpha ^{\prime }+2\beta ^{\prime }=0$. An association of two equations $%
1-\alpha ^{\prime }+\beta ^{\prime }=0$ and $4-3\alpha ^{\prime }+2\beta
^{\prime }=0$ yields, 
\begin{equation}
\alpha ^{\prime }=2,\beta ^{\prime }=1.\text{i.e., }a^{\prime }\left(
T_{C}\right) =2a\left( T_{C}\right) /T_{C},b^{\prime }\left( T_{C}\right)
=b\left( T_{C}\right) /T_{C}.
\end{equation}%
With these values, we find that not only the critical point is locally flat,
but also $a\left( T\right) $ and $b\left( T\right) $ are, accurate up to the
first order of $(t^{\ast }-1)$, 
\begin{subequations}
\begin{eqnarray}
a\left( T\right)  &\approx &a\left( T_{C}\right) +2a\left( T_{C}\right)
(t^{\ast }-1)=-a\left( T_{C}\right) +2a\left( T_{C}\right) t^{\ast }, \\
b\left( T\right)  &\approx &b\left( T_{C}\right) +b\left( T_{C}\right)
(t^{\ast }-1)=b\left( T_{C}\right) t^{\ast }.
\end{eqnarray}%
When $t^{\ast }\approx 1$, i.e., $T\approx T_{C}$, $a\left( T\right) \approx
a\left( T_{C}\right) $ and $b\left( T\right) \approx $ $b\left( T_{C}\right) 
$, the usual form of vdW EoS is assumed. It is important to note that, from
two relations above, the usual vdW EoS is valid when the thermodynamic
states are very close to the critical point, and $a\left( T\right) $ and $%
b\left( T\right) $ are also solely determined by $a\left( T_{C}\right) $ and 
$b\left( T_{C}\right) $.

\section{Conclusions}

Differential geometry is powerful tool to reveal the intrinsic nature of the
curved surface, and it is advantageous to analyze the critical point on the
EoS surface. On the tangential plane of the critical point, the existence of
limit $\left( \partial p/\partial V\right) _{T}=0$ requires the isothermal
process. However, the essence of the critical point is its independence of
whatever process is taken, and of detailed structure of matters, etc. We can
therefore assume $\left( \partial p/\partial T\right) _{V}=0$ and $K_{C}=0$
at the critical point on the EoS surface. 

The vdW EoS is the simplest one to understand the liquid-gas transition.
Since the vdW parameters $a$ and $b$ are constant, the Gaussian curvature is
negative definite; and there is no distinction between vdW EoS and the ideal
gas EoS. According to our assumptions, the vdW EoS is slightly modified or
extended such that the vdW parameters $a$ and $b$ vary with temperature,
allowing for presence of the zero Gaussian curvature at the critical point.
Our approach sheds light on understanding the theoretical problem how the
vdW parameters depend on the temperature.

\begin{acknowledgments}
This work is financially supported by National Natural Science Foundation of
China under Grant No. 11675051.
\end{acknowledgments}

\end{subequations}

\end{document}